\documentstyle[psfig]{mn}

\let\jnlst=\rm
\def\refjnl#1{{\jnlst#1}}

\def\aj{\refjnl{AJ}}                   
\def\araa{\refjnl{ARA\&A}}             
\def\apj{\refjnl{ApJ}}                 
\def\apjl{\refjnl{ApJ}}                
\def\apjs{\refjnl{ApJS}}               
\def\aap{\refjnl{A\&A}}                
\def\mnras{\refjnl{MNRAS}}             
\def\pasp{\refjnl{PASP}}               

\let\citeANP=\cite

\title{The colour magnitude relation for galaxies in the Coma cluster.}
\author[A. I. Terlevich et al.]
  {A.~I.~Terlevich$^1$,~N.~Caldwell$^2$, ~R.~G.~Bower$^3$.\\
  $^1$University of Birmingham, Edgbaston, Birmingham. B15 2TT.\\
  $^2$F.L. Whipple Observatory, Smithsonian Inst., PO box 97, Amado, Az 85645, USA.\\
  $^3$University of Durham, South Rd. Durham. DH1 3LE.}
\date{Accepted 6th June 2001}
\pagerange{\pageref{firstpage}--\pageref{lastpage}}
\pubyear{2001}

\begin{document}
\label{firstpage}

\maketitle

\begin{abstract}

We present a new photometric catalogue of the Coma galaxy
cluster in the Johnson U- and V- bands. We cover an area of 3360${\rm
arcmin}^2$ of sky, to a depth of $V=20$ mag in a 13 arcsec diameter
aperture, and produce  magnitudes for $\sim 1400$ extended objects in
metric apertures from $8.8$ to $26$arcsec diameters.  The mean
internal RMS scatter in the photometry is $0.014$ mag in V, and
$0.026$ mag in U, for $V_{13} < 17$ mag.

We place new limits on the levels of scatter in the colour--magnitude
relation (CMR) in the Coma cluster, and investigate how the slope and
scatter of the CMR depends on galaxy morphology, luminosity and
position within the cluster. As expected, the lowest levels of scatter
are found in the elliptical galaxies, while the late type galaxies
have the highest numbers of galaxies bluewards of the CMR. We
investigate whether the slope of the CMR is an artifact of colour
gradients within galaxies and, show that it persists when the colours
are measured within a diameter that scales with galaxy size.  Looking
at the environmental dependence of the CMR, we find a trend of
systematically bluer galaxy colours with increasing projected radius
from the center of the cluster. Surprisingly, this is accompanied by a
{\it decreased} scatter of the CMR. We investigate whether this
gradient could be due to dust in the cluster potential, however the
reddening required would produce too large a scatter in the colours of
the central galaxies. The gradient appears to be better reproduced by
a gradient in the mean galactic ages with projected radius.
\end{abstract}

\section{Introduction}

The progressive reddening of the integrated colours of elliptical
galaxies with increasing luminosity is known as the colour-magnitude
relation (CMR) (Faber~1973; Visvanathan \& Sandage~1977; Frogel
et~al.~1978; Persson et~al.~1979; Bower, Lucey \& Ellis~1992a,1992b
[BLE92a,BLE92b]. Despite being a far simpler relation than the
Fundamental Plane
\cite{Dressler87,Djorgovski87,Bender92,SagliaBD93,Jorgensen93,Pahre95},
it nonetheless has similarly small levels of scatter. Traditionally,
the slope seen in the CMR has been attributed to a mass--metallicity
sequence \cite{Dressler84,Vader86}, with the massive galaxies being
more metal rich, and thus redder, than the less massive ones.  This
tendency can naturally be explained by a supernova--driven wind model
\cite{Larson74,ArimotoYoshii87}, in which more massive galaxies can
retain their supernova ejecta for longer than can smaller galaxies,
thus being able to process a larger fraction of their gas before it is
expelled from the galaxy. In hierarchical models of galaxy formation
(e.g., Kauffmann \& Charlot, 1998), the CMR can be reproduced because
metals are expelled from low mass galaxies as part of the feed-back
process.  Studies of the CMR in high redshift clusters find a
ridge-line slope comparable to that of the local clusters;
furthermore, there is no sign of a change in the range of magnitude
over which the CMR may be traced
\cite{EllisMORPH97,StanfordED98,KodamaArimoto97,Kodama97,KodamaABA98}.
These studies all point towards the galaxies which make up the CMR in
the cores of rich clusters being primarily constituted of uniformly
old stellar populations. Given this metallicity driven interpretation
of the CMR, its low levels of scatter in cluster cores implies that
the galaxies are made up from uniformly old stellar populations (BLE92b;
Bower, Kodama \& Terlevich~1998).  Even small variations in the ages
of the galaxies would lead to unacceptable levels of scatter in young
stellar populations, whereas old stellar populations have a much
smaller age dependency in their colours.

Despite the uniformity of the CMR between the Coma and Virgo cluster
cores (BLE92b), studies of the CMR in Hickson compact groups
\cite{Zepf91} show increased scatter. Similarly, studies of field
ellipticals \cite{LarsonTC80} indicate that the scatter in the CMR
could depend on environment. However these group and field galaxy
samples contain data from many disparate sources, therefore there
might be an added source of scatter from matching the various
photometric datasets onto a single photometric system. These
additional sources of scatter can take the form of uncertainties in
K-corrections for galaxies at different redshifts, or biases
introduced due to sampling only the brighter end of the luminosity
function at higher redshift in a magnitude limited sample. Despite
this, at least in the sample of \citeANP{LarsonTC80}, the extra
scatter seems too large to be accounted by increased observational
uncertainties alone.  Kodama et al.\ (1999) analysed the CMR in the
Hubble Deep Field North.  Again they found an increase in the CMR
scatter, and a possible indication that the slope of the CMR is
flatter at high redshift in the field environment.

Further evidence for an environmental dependence of the CMR comes from
studies of spectral line indices. Broadband colours are notoriously
inefficient at separating the effects of age and metallicity on a
stellar population, a degeneracy neatly summarised by Worthey (1994)
in his `$2/3$' law ($\Delta[Fe/H] \sim {2\over3} \Delta log(t)$). This
has led to studies of the stellar populations of early type galaxies
in the cores of clusters using spectral line indices chosen to break
this degeneracy and disentangle the effects of age and
metallicity. Such studies (e.g. Mehlert et al.~1998; Kuntschner~1998;
Kuntschner \& Davies~1998) show that the CMR is driven by metallicity
variations with galaxy luminosity, rather than age. A direct
comparison of colours and line-index methods by Terlevich et al.\
(1999) showed that the two approaches have comparable sensitivity.

The key question then is to determine how the uniform CMR seen in
cluster cores transforms into the less tight CMR of field galaxies.
In order to investigate how the CMR varies across the Coma cluster,
between different regions and between different galaxy morphological
types, we have undertaken a survey of $(U-V)$ colours covering almost
one square degree of the Coma cluster. Although the colours of the
galaxies in Coma have been studied before, both in a wide area
(e.g. Dressler et~al.~(1980); Godwin et~al.~(1983) [GMP]) and with
high precision U and V band CCD data (BLE92a),
the present study is unmatched in area and sensitivity to variations
in stellar populations.  We chose to use the Johnson \cite{Johnson53}
U and V filters because they straddle the $4000$\AA\ break in the
spectra of galaxies at low redshift, and are thus very sensitive to
the ages of the stellar populations. They are especially sensitive to
recent bursts of star formation (e.g. Worthey~1994; Charlot \&
Silk~1994)

The present study extends the photometry of BLE92a to a complete
galaxy sample covering approximately four times the area (still
centered on the core), and reaches to fainter limiting magnitudes.
The extra coverage and depth will enable us to obtain colours for the
abnormal spectrum `E+A' galaxies of Caldwell~et~al.~(1993;~1996), many
of which are to be found in the South West corner of the cluster
around a group of galaxies dynamically associated with NGC4839
\cite{Baier84,Escalera92,CollessDunn96}.  Significant advances in
detector technology since the work of BLE92a, allows us to use much
larger CCDs with greater U-band sensitivity. In order to cover the
required area, we took tiled images giving us continuous coverage of
the cluster. In contrast, BLE92a targeted individual galaxies. Because
of the continuous coverage, our sample of galaxies is more complete
than that of BLE92a, including all of the GMP galaxies within our area
of sky and with significantly higher precision than the GMP data.

\section{Observations}

\begin{figure}
\centering
\centerline{\psfig{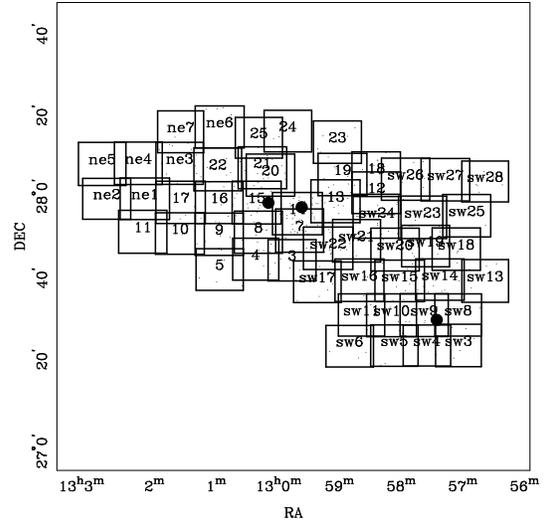}}
\caption{The distribution of observed images across Coma. All observed
galaxies with $V_{16} > 18$ are shown as a dot. The large dots show,
from left to tight, the positions of NGC4889, NGC4874 and NGC4839
respectively. The dynamical center of the Coma cluster is somewhere
between NGC4889 and NGC4874, while the dynamical center of the
substructure in the SW corner is NGC4839.}
\label{fig:framenames}
\end{figure}
  The observing runs which provided data for use in this project are
summarised in table \ref{obs}. U and V-band observations were obtained
in two successive years at the SAO 1.2m on Mt Hopkins, Arizona. The
detector used was a thinned, back-side illuminated, AR coated
$2048\times2048$ Loral CCD in $2\times 2$ binning mode, giving us a 10
arcmin ($192h^{-1}kpc$) field of view with $0.63$ arcsecond pixels. The
quantum efficiency of the CCD stays high and almost constant right
across the U band, giving effective filter responses which approximate
the standard shape. The same setup was used for all observations to
maintain a common photometric system for the whole dataset. The
average integration times used were 400s in the V-band and 45min in
the U-band and the median V-band seeing achieved throughout the run was 2.2
arcsec FWHM (see appendix A).

  The observations cover a continuous region encompassing the South
West group around NGC4839, the central parts around NGC4874 and
NGC4889 and also a large amount of the North East of the cluster (see
figure \ref{fig:framenames}). The observations were also designed to
cover all of the Caldwell~et~al.~(1993; 1996) abnormal spectra
galaxies, however inclement weather meant that some were missed
towards the extreme SW and NE. The presence of a seventh magnitude
star just north of the center was also avoided as scattered light here
makes data reduction difficult. To reduce any systematic differences
in the photometry between parts of the cluster, observations of the
central and SW regions were interleaved during the observing runs.

 During the night, immediately after dusk and before twilight,
standard stars from Landolt~(1992) were observed over a wide range in
airmass. Care was taken to ensure the colours of the stars matched
those of our galaxies, typically $0\le(U-V)\le2$. With our large field
of view it was possible to observe many standards simultaneously. To
have additional checks on the overall homogeneity of the final
photometry, we used large overlaps of $\sim 1$~arcmin between the
actual Coma cluster images, ensuring that the objects in this overlap
region were observed in both images. We also interleaved snapshots
(300s and 100s for U and V respectively) of the central parts of the
cluster, thus using the galaxies there as `standard' galaxies.  This
is particularly important for the U band, as the spectral energy
distribution of the standard stars is different from that of the early
type galaxies.

\begin{table}
\caption{Summary of observing time.}
\centering
\centerline{
\begin{tabular}{lll}
Dates & Observer(s) & Usable Nights\\
\hline
20--21 March, 1996 & Caldwell & 1.5 \\
11--14 April, 1996 & Caldwell \& Terlevich & 4 \\
9--11 May, 1996 & Caldwell & 4 \\
1--5 April, 1997 & Caldwell \& Terlevich & 1.5  \\
\hline
\end{tabular}
}
\label{obs}
\end{table}

\section{Data reduction}

The images were reduced using the standard methods in the IRAF
package. The CCD used had a number of cosmetic defects, so in the
subsequent reduction procedures, we mark objects within 5 pixels of a
defect as suspect.

\subsection{Galaxy identification, photometry and astrometry}
\label{sec:photometry}
  Lists of candidate objects were produced using the Sextractor1.2b10
program \cite{Bertin96} from the V-band images. Sextractor was also
used to differentiate the extended sources from the stellar sources in
the images. All objects within 15 pixels ($9.45{\rm arcsec}$) of a CCD
edge were rejected. In order to avoid problems matching galaxies on
the U-band frames, galaxies fainter than $V = 20$ were not carried 
forward for further analysis. This gives rise to a sharp cutoff in the
magnitude distribution of our catalogue.

The list of positions for the galaxies in each image was used by the
IRAF {\tt phot} package to generate fixed aperture magnitudes in
$8.8\rm{arcsec} - 30\rm{arcsec}$ diameter apertures. The sky level was
measured in annuli with inner radii of 50 arcsec. For the fixed
photometric apertures with radii less than 20arcsec, we also measured
the sky level at 25arcsec, and we give both values in the final data
table.  The magnitudes measured were then corrected for the varying
seeing between the frames as described in Appendix~\ref{sec:seeing}.
Astrometry for the frames was calculated using the HST guide star
catalogue as a source of reference stars. The RMS scatter in our
astrometry is approximately $1$~arcsec. We have not corrected the data
for geometric distortions in our detector, as they introduce
photometric errors of less than $1\%$ between galaxies at the centre
and galaxies at the edge of the detector.

Observations of standard stars left residuals at the $0.03{\rm mag}$
level in the U-band, once a drift in the zero point had been
corrected. The drift (in all cases less than 0.2 mag per night) is
also evident when examining the overlap regions between images and the
repeated observations of `standard' Coma fields. As these offsets were
so readily measurable, we used them to improve the zero points of the
individual images in order to ensure that the whole cluster is on a
consistent photometric system for all the observations. The method we
used to generate this system is very similar to that used by
Maddox~et~al.~(1990), and is described in detail in Appendix~B. The
absolute calibration of this system was set to agree with the
published U and V band photometry of BLE92a.

\subsection{Quantifying the photometric errors}
\label{sec:photerrs}

\begin{figure}
\centering
\centerline{\psfig{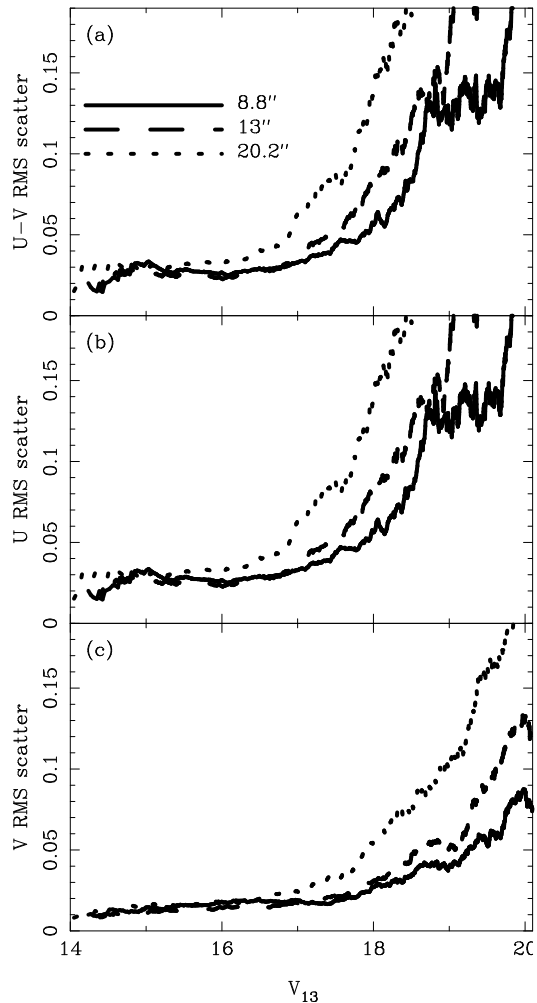}}
\caption{ In the three panels, we plot the observational errors as a
function of $V_{13}$ magnitude, from repeated measurements of galaxies
in overlapping regions.  The $(U-V)$ colour errors are shown in panel
(a). Panels (b) and (c) show the errors for the U and V
photometry. The three lines show a running bi-weight scatter (see
section~\ref{sec:photerrs}) measured through circular apertures of
$8.8\arcsec$,$13\arcsec$ and $20.2\arcsec$ diameter. A bin size of 60
observations was used in calculating the running mean, reducing to 6
at the bright extreme of the plot.  Table~\protect\ref{tab:photscat}
lists the mean RMS scatter down to $V_{13} = 17$mag for all of our
apertures. The increasing levels of scatter with aperture size is entirely consistent with the increasing
contribution of the sky to the noise with aperture radius. }
\label{fig:UVrunphotscat}
\end{figure}
\begin{figure*}
\centering
\centerline{\psfig{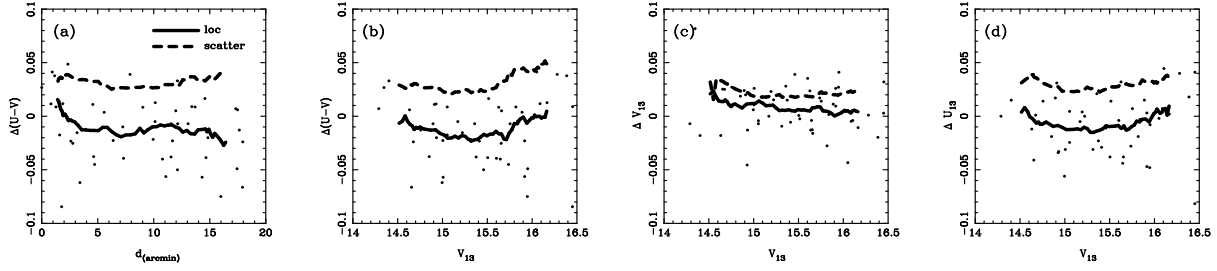}}
\caption{ The four panels show the behaviour of the residuals between
our $13$arcsec diameter aperture photometry and that of BLE92a. In all
cases, the residuals are calculated by subtracting the BLE92a data from
ours, for example, a negative $\Delta(U-V)$ indicates that our colour
for a galaxy is bluer than the BLE92a colour. The Solid and dashed lines
show running biweight location and scatter indicators.  Panel (a)
shows the difference between the $(U-V)_{13}$ colours obtained in this
paper and those of BLE92a, as a function of distance from NGC4874.  Panel
(b) shows how our colours compare to the BLE92a colours as a function of
luminosity.  Panels (c) and (d) show how our U and V band magnitudes
compared with those of BLE92a.}
\label{fig:UVRGBradscat}
\end{figure*}
\begin{figure}
\centering
\centerline{\psfig{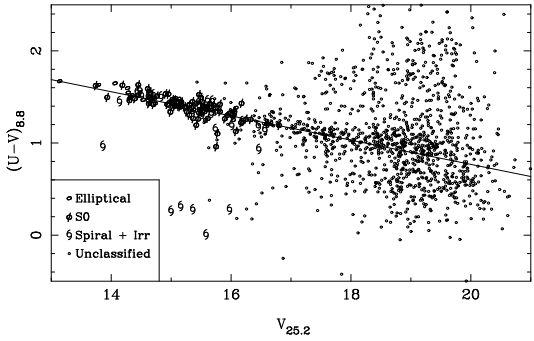}}
\caption{ The $(U,V)$ colour magnitude relation for all objects
detected for which Sextractor gives a $\rm{CLASS\_STAR}\le 0.2$. The
$V$ are taken from the $25.2''$ diameter aperture, as it is more
complete than the $26''$ aperture. In order to increase the signal to
noise in the colour term, the $(U-V)$ are taken from the smallest
aperture, the $8.8''$ diameter aperture. The symbols represent the
morphological types of the galaxies from Andreon~(1996;~1997). The
small open symbols have no morphological information. The line is a
biweight fit (see section~\ref{sec:photerrs}) to the data, and follows
the ridge-line of the CMR which can be seen to extend down to
$m_{V25.2}\sim19.5$. }
\label{fig:CMallobj}
\end{figure}
\begin{table}
\caption{This table lists the RMS errors in our photometry for all of
our apertures. It was generated using the repeated observations of
both stars and galaxies. mostly in the regions of overlap between
images. Figure~\protect\ref{fig:UVrunphotscat} shows how the RMS
errors vary with magnitude for the $8.8$,$13$ and $20.2$ arcsec
diameter apertures. The RMS internal scatter quoted by BLE92a for their
$13''$ photometry, which only reaches a magnitude of $V_{13}=16.5$ is
$0.025$ and $0.015$ mag for U and V respectively.}
\centering{
\begin{tabular}{cccc}
Aperture diameter & \multicolumn{3}{c}{\hrulefill RMS scatter \hrulefill}\\
(arcsec) & $U$ & $V$ & $(U-V)$ \\
\hline
8.8  & 0.02303 & 0.01619 & 0.02829\\
12.6 & 0.02124 & 0.01364 & 0.02532\\
13   & 0.02258 & 0.01352 & 0.0264\\
16   & 0.0277  & 0.01478 & 0.03145\\  
20.2 & 0.03036 & 0.01727 & 0.035\\
25.2 & 0.0359  & 0.02076 & 0.04154\\
26   & 0.04805 & 0.02706 & 0.05528\\
\hline
\end{tabular}}
\label{tab:photscat}
\end{table}   

 We can use the large number of multiply observed objects (mostly from
the large overlaps between images) to constrain our observational
errors. Table \ref{tab:photscat} lists the mean RMS scatter for
all objects with $V_{13} < 17$ mag in all of our apertures, and Figure
\ref{fig:UVrunphotscat} shows how the scatter in the observations
of multiply observed objects increases with magnitude for the
$8.8$,$13$ and $20.2$arcsec diameter apertures. The scatter was
computed using a running biweight scatter indicator \cite{Beers90}. We
use the biweight scatter indicator throughout this paper due to its
robustness and resistance in the case of non Gaussian
distributions. Additionally, we perform all regression analysis by
minimising both the biweight scatter and location (mean) of the
residuals to the fit. This provides an efficient way of performing
fits to data which in the case of the CMR residuals, have a non
Gaussian distribution, and often a large tail.

 The levels of scatter in our $13$arcsec apertures are the same as
those quoted by BLE92a ($0.025$ and $0.015$ mag for U and V respectively),
and they stay constantly low down to $V_{13} \sim 17$ mag. It should be
noted that although table~\ref{tab:photscat} shows that the
$12.6''$ aperture has the lowest RMS scatter in both U and V bands,
figure~\ref{fig:UVrunphotscat} shows that fainter than $V_{13}\sim
17$, it is quickly overtaken by the $8.8''$ aperture, which is better
suited to the smaller sizes of the fainter galaxies.

As an independent check of our calibration, we have compared our
photometry directly with that of BLE92a (see
fig. \ref{fig:UVRGBradscat}).  The scatter between our colours and
theirs is 0.034 mag, while the scatter between our photometry and
theirs is $0.022$ and $0.032$ mag for the V and U-bands
respectively. This is almost exactly what we expect simply by adding
the rms internal scatters of our data and theirs in quadrature.
Equally important, given the method used to obtain a uniform
photometric system for our data, is the fact that the mean colour
difference does not vary as a function of distance from NGC4874
(figure~\ref{fig:UVRGBradscat}, panel a).

\subsection{The Photometric Diameter $D_V$}

The photometric diameter parameter, $D_V$ is equivalent to the $D_n$
parameter used by Dressler et~al.~(1987), but is based on V-band
photometry. We use the definition of $D_V$ given in Lucey
et~al.~(1991), viz. ${D}_V$ is the photometric diameter (in arcsec)
which encloses an area of average surface brightness of $19.80\
\rm{mag\ arcsec}^{-2}$. Like those of Lucey et~al., our $D_V$ include
a $(1+z)^4$ cosmological correction. As the $D_V$ are (mostly)
calculated using interpolation, they are very accurate. In order to
measure $D_V < 8.8$arcsec (our smallest photometric aperture), we also
need to perform some extrapolation, and we use $D_V$s only down to
$4$arcsec to keep this to a stable minimum. The function we use for
both interpolating and extrapolating is a simple $R^{1/4}$ profile,
which is fitted to the seeing corrected aperture magnitudes. In
addition to rejecting $D_V$s smaller than $4$arcsec, we also do not
attempt to calculate $D_V$s for galaxies where the function fit was
poor, or where extrapolation of the data to larger radius is
necessary. Comparison with the independent $D_V$ values given by Lucey
et~al.~(1991) for the same galaxies shows the uncertainty to be better
than $0.007dex$.

\subsection{Catalogue}

An extract of the photometric catalogues are presented in Table 3. The
full version, which is only available electronically, contains U and
V-band photometry for all the apertures listed in table
\ref{tab:photscat}.

\begin{table*}
\caption{A small extract of the photometric catalogue. Due to its
size, The full catalogue is only available electronically.}
\centering{
\begin{tabular}{cccccccc}
 GMP & RA J2000 & DEC J2000 & log$_{10} (D_V)$ & $V_{13}$ & $U_{13}$ & $V_{16}$ & $U_{16}$\\
\hline
 1807 & 13:01:50.2 & 27:53:36.2 & 0.943 & 15.322 & 16.727 & 15.161 & 16.570\\
 1853 & 13:01:47.0 & 28:05:41.5 & 1.055 & 14.866 & 16.317 & 14.704 & 16.167\\
 1885 & 13:01:44.1 & 28:12:51.4 & 0.628 & 16.747 & 17.957 & 16.630 & 17.845\\
 2000 & 13:01:31.8 & 27:50:50.9 & 1.112 & 14.636 & 16.105 & 14.456 & 15.924\\
 2048 & 13:01:27.2 & 27:59:56.8 & 0.821 & 16.039 & 17.435 & 15.938 & 17.338\\
 2059 & 13:01:26.2 & 27:53:09.9 & 1.070 & 14.774 & 16.288 & 14.565 & 16.099\\
\hline
 \end{tabular}
}
\label{tab:ExampleCat}
\end{table*}

\section{Analysis technique}
\label{sec:analysis}

The first step in our analysis is to establish a criterion for cluster
membership. Redshifts are available for all galaxies brighter than
$V_{13} = 15.7$. For the fainter galaxies, $96\%$ of galaxies with
$V_{13} < 16$, $89\%$ of $V_{13} < 17$ galaxies, and $71\%$ of $V_{13}
< 18$ galaxies have redshifts. The velocities were obtained from the
NASA/IPAC Extragalactic Database\footnote{The NASA/IPAC Extragalactic
Database (NED) is operated by the Jet Propulsion Laboratory,
California Institute of Technology, under contract with the National
Aeronautics and Space Administration.}  (NED). Most of the velocities
can be attributed to Colless \& Dunn~(1996).  Galaxies with
recessional velocities between $4000 kms^{-1}$ and $10000 kms^{-1}$
were taken as confirmed members of the Coma cluster. From this sample
we then removed galaxies with `bad' photometry. Galaxies were deemed
bad, and thus rejected if
\begin{itemize}
\item{ Emission from a nearby object entered the $13''$ diameter
aperture. }
\item{ One of the CCD bad columns passes through the galaxy.}
\item{ A cosmic ray was removed from part of the galaxy in the V image
(Cosmic rays in the U images were less of a problem due to the image
being composed of multiple exposures).}
\end{itemize}

The morphological classification of galaxies is taken from Andreon
et~al.~(1996; 1997) which gives morphologies for all $V_{13} < 15.7$
galaxies, but the morphologies don't go as faint as the redshifts.
$90\%$ of $V_{13} < 16$ galaxies, $68\%$ of $V_{13} < 17$ galaxies and
$38\%$ of $V_{13} < 18$ galaxies have morphological information.  The
elliptical and late types are uniformly distributed throughout this
magnitude range, however the proportion of S0 galaxies relative to the
total number, has a sharp peak at $V_{13}=16$mag, where $68\%$ of the
galaxies are S0s.  The different symbols in figures
\ref{fig:CMallobj},\ref{fig:CMall}, \ref{fig:datasets1} and \ref{fig:dvcmr}
correspond to the broad morphological type of the galaxy. The actual
morphological types used in these broad classifications are shown in
table \ref{tab:MorphNumbers}, together with the frequency of each
type.

\begin{table*}
\caption{The morphological classes used in this paper, combined into
the broad categories of late-type, S0, and early type. Throughout the
paper, no distinction shall be made between disky, boxy and undefined
ellipticals (diE, boE, unE). They shall simply be referred to as
ellipticals.}
\centerline{
\begin{tabular}{llc}
Category& Morphological type  & Number of galaxies\\
\hline
late-type& Sp,S,SBa, Sa, SA0/a, SB0/a, SAB0/a, SAB0p& 32\\
S0 type& SA0, SAB0, SB0, diE/SA0, diE/SAB0, unE/SA0&   71\\
early type&  Epec, boE, diE, unE & 26\\
No morph &  &148\\
\hline
\end{tabular}}
\label{tab:MorphNumbers}
\end{table*}

Regression analysis of the colour magnitude relation was performed
using the biweight estimator. Our aim is to distinguish the ridgeline
of the CMR, and we do not want to be unduly influenced by exact 
position of blue outliers. The biweight is a robust,
resistant and efficient location and scale indicator, as is apparent
from table \ref{tab:regress}. We compared the results of 
the biweight technique to that of using a biweight technique in
conjunction with a $3\sigma$ clipping of the dataset, however
the difference was within the $1\sigma$ error estimate of the
un-clipped method. 

The errors in the best fit relation were calculated by bootstrap
resampling of the data.  The observational uncertainties in the colour
as a function of galaxy luminosity ($O_c(L)$) are well known for this
dataset (see fig. \ref{fig:UVrunphotscat}), so in order to make an
estimate of how much of the measured scatter in the CMR is due to
observational errors, we defined a mean observational colour scatter thus,
\[
\bar{O}_c = {\sum_{i=1}^N O_c(L_i)\over N}
\]
where the $L_i$ are the luminosities of the galaxies in the dataset, and
$N$ is the number of galaxies. Using this value for the observational
errors, a value for the intrinsic scatter in the CMR can be
calculated.
\begin{equation}
\label{eqn:quadadd}
I =  \sqrt{\sigma^2 - \bar{O}_c^2}
\end{equation}
where $\sigma$ is the observed scatter of the CMR.

Due to the fact that we are measuring the CMR scatter using a limited
sampling of the underlying distribution, some uncertainty is
introduced. This can make the measured scatter ($\sigma$) smaller than
the observational errors ($\bar O_c$).  In such a cases, we show the
intrinsic scatter as zero, but we can still show an upper limit to the
intrinsic scatter from the bootstrap limits.

\section{Environmental and morphological variations}
\label{sec:CMRs}
\begin{table*}
\caption{Selection criterea for each dataset. All datasets are a subset of dataset 1.}
\begin{tabular}{cp{13cm}}
Dataset &Description\\
\hline
1 & Confirmed members:  After rejecting the `bad' galaxies (see
section~\ref{sec:analysis}), we classify all galaxies from $4000
 kms^{-1}$ to $10000 kms^{-1}$ as members of the cluster. The
 velocities were obtained from the NASA/IPAC Extragalactic Database 
(NED). Most of the velocities can be attributed to Colless \& Dunn~(1996).\\

2 & Confirmed members with $V_{13} < 17$: A subsample of confirmed
members (dataset 1) with the faint tail cut off at the point where the
measurement errors in the colours starts to increase (see
figure~\protect\ref{fig:UVrunphotscat}).\\

3& All with morphology: All member galaxies with a morphology from
Andreon et~al.~(1996; 1997) (see
table~\protect\ref{tab:MorphNumbers}).\\

4& E\&S0 morphology: All member galaxies with an elliptical or S0
morphological type (see table~\protect\ref{tab:MorphNumbers}).\\

5& S0 morphology: All member galaxies with S0 morphology (see
table~\protect\ref{tab:MorphNumbers}).\\

6& Elliptical morphology:  All member galaxies with Elliptical morphology (see
table~\protect\ref{tab:MorphNumbers}).\\

7 & Late type morphologies: All member galaxies with late type
(spiral and irregular) morphology (see
table~\protect\ref{tab:MorphNumbers}).\\

8 & E\&S0 center: Early type galaxies closer to NGC4874 than NGC4839.\\

9 & E\&S0 SW: Early type galaxies closer to NGC4839 than NGC4874.\\

10 & E\&S0 inner: Early  type galaxies within $15'$ of either NGC4874
or NGC4839.\\

11 & E\&S0 outer: Early type galaxies further than $15'$ from both
NGC4874 and NGC4839.\\
12 & E\&S0 bright: The bright half of dataset 4\\
13 & E\&S0 faint: The faint half of dataset 4\\
14 & Members bright: The bright half of dataset 1\\
15 & Members faint: The faint half of dataset 1\\
\hline
\end{tabular}
\label{tab:Cuts}
\end{table*}
\begin{figure}
\centering
\centerline{\psfig{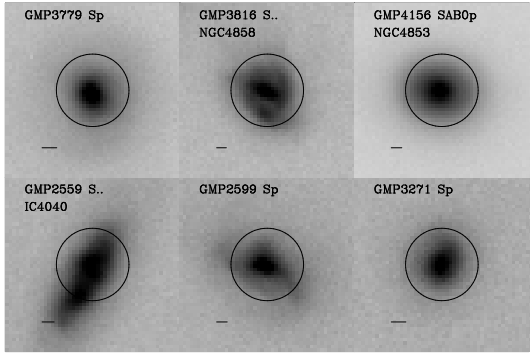}}
\caption{V band images of the late type galaxies
which lie blue-wards of the measured CMR for late type galaxies (dataset
7) by more than $3\sigma$.  The four galaxies to the left of the plot
are the four (out of a total of seven) Bothun \& Dressler (1986)
sample of blue disk galaxies in our surveyed environment. Out of these
four, NGC4856 is the only galaxy not to have H$\alpha$ in
emission. Bothun \& Dressler conclude that these galaxies are
undergoing a short burst of star formation activity. Additionally,
NGC4853 is a post starburst galaxy identified by
Caldwell~et~al.~(1993), and is included here with the late types, as
it has peculiar asymmetries in its light profile (Andreon et~al.~
1997). The circle in each figure represents the $13''$ diameter
aperture used in the measurements of both the colours and the
magnitudes for the results in table \protect\ref{tab:regress}. The
line in the lower left of each panel shows the width of the seeing
disk in each image.}
\label{fig:Sp_blue}
\end{figure}
\begin{figure}
\centering
\centerline{\psfig{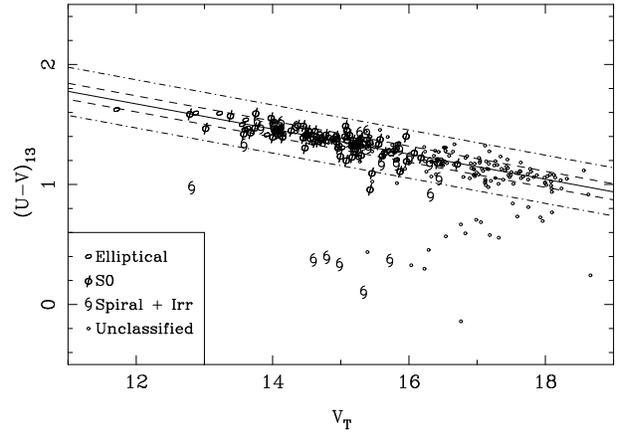}}
\caption{ The colour magnitude relation for all galaxies with
recession velocity within $3000kms^{-1}$ of $7000Kms^{-1}$ (dataset
1). The magnitudes are estimated total V magnitudes ($V_T$). Different
colour symbols represent different morphological types as determined
from Andreon et~al.~(1996;~1997).  The solid line shows the best fit
to the data using the biweight minimisation technique (see text). The
solid line is a best fit to all the data points. The dashed lines show
the $1\sigma$ and $3\sigma$ scatter.}
\label{fig:CMall}
\end{figure}

Figure~\ref{fig:CMallobj} shows the colour magnitude relation for
every extended object in the photometric catalogue. Many of the
objects shown will not be members of the Coma cluster, yet despite
this, the CMR is clearly visible down to $V_{25.2} = 19$ mag . We are
interested in measuring changes in the CMR, such as in its scatter or
slope, in different parts of the cluster, in different subsets of
galaxy morphology, and for different luminosities. We therefore
concentrate solely on those galaxies identified as members of the
cluster from their recessional velocity. Using the recessional
velocity avoids the need for statistical background subtraction. We
have used the galaxies' properties, such as morphology, luminosity and
position, to define 14 subsets of these 275 member galaxies. The
subsets are defined in table~\ref{tab:Cuts}.

Figure \ref{fig:CMall} shows the CMR for all confirmed cluster members
(dataset 1). The CMR is made up of galaxies of differing morphological
types, and from every part of the cluster, yet it extends for almost
five magnitudes without deviating from a straight line. Figure
\ref{fig:CMallobj} shows that the CMR actually extends fainter than
this in our data, but we have no redshifts for these faint
galaxies. Secker et~al.~(1997) have shown that the $(B,B-R)$ CMR in
dwarf ellipticals in Coma, actually continues down to at least $B\sim
21.5$. Another important aspect of figure \ref{fig:CMall} is in the
direction of scatter. There is almost no scatter red-ward of the CMR
ridge line, even at the faint end where the observational errors are
greatest, there is however significant blue-ward scatter, most of
which is due to the late type population.

In the following sections, we investigate the properties of the CMR in
each of the datasets (see figure~\ref{fig:datasets1}. We use the
techniques described in section~\ref{sec:analysis} to ascertain the
scatter about the main ridge line of the CMR, as well as the scatter
in the total sample. Throughout the rest of the paper we use the
$13$arcsec diameter aperture magnitudes and colours (due to their low
photometric errors) or colours measured within the $D_V$ diameter (in
order to define colours within an aperture that scales in galaxy
size).

\begin{figure*}
\centering
\centerline{\psfig{figure=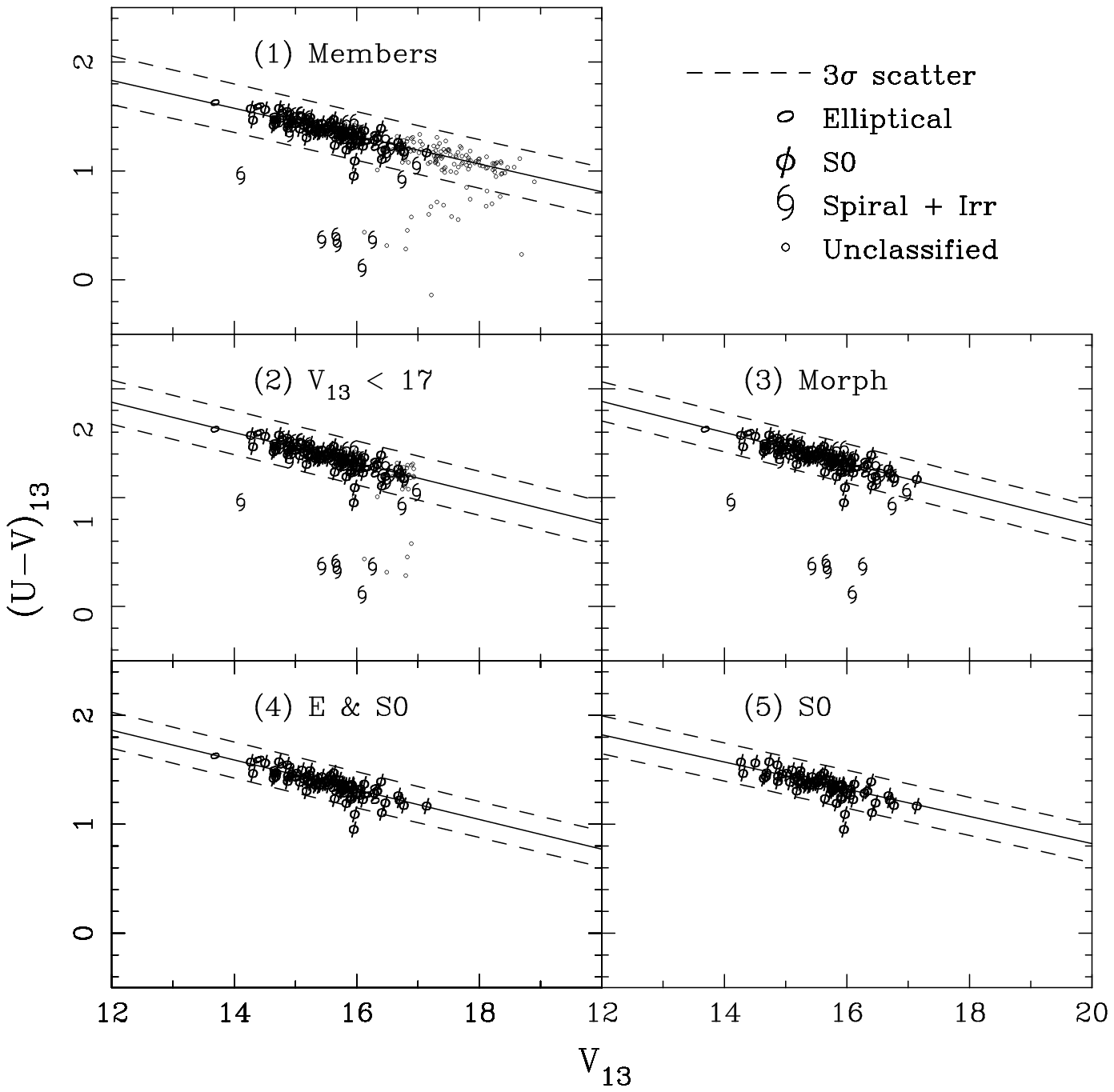,width=14cm}}
\caption{ This figure (continued on pages \protect\pageref{fig:datasets2}
and \protect\pageref{fig:datasets3}) shows the CMRs for each of the
datasets defined in the main text. The symbols are the same as those
used in figure \protect\ref{fig:CMall}. Dashed lines
represent the $3\sigma$ scatter of the galaxies about the best fit
line for the full dataset.}
\label{fig:datasets1}
\end{figure*}
\begin{figure*}
\centering
\centerline{\psfig{figure=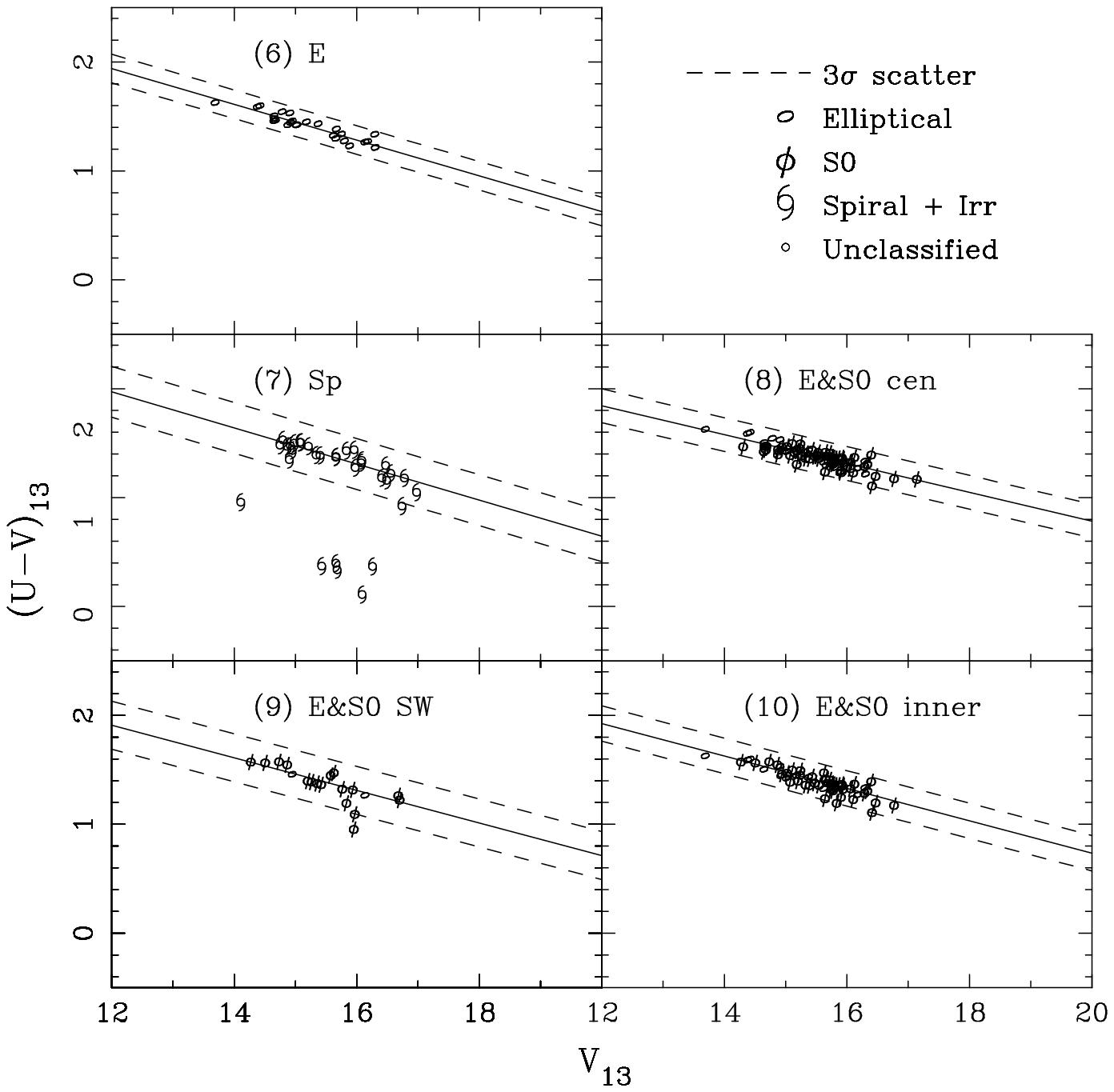,width=14cm}}
\contcaption{}
\label{fig:datasets2}
\end{figure*}
\begin{figure*}
\centering
\centerline{\psfig{figure=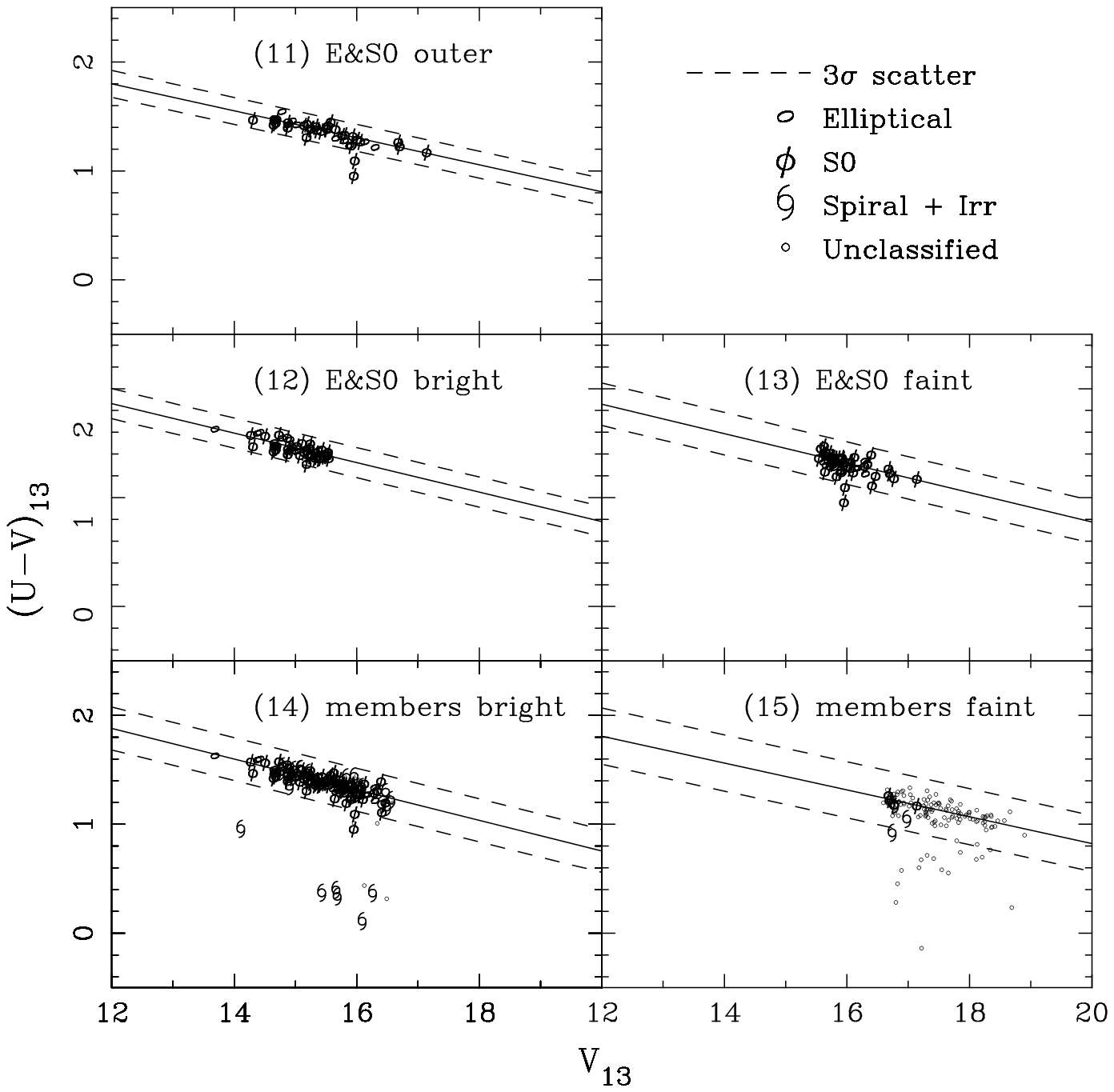,width=14cm}}
\contcaption{}
\label{fig:datasets3}
\end{figure*}
\begin{table*}
\caption{Results of regression analysis on the $13arcsec$ diameter
aperture magnitudes. The errors quoted for the slope,
intercept and observed scatter, are $1\sigma$ bootstrap errors. The
mean observational. scatter ($\bar O_c$) is used in calculating the intrinsic
scatter (see main text).}
\begin{tabular}{lcccccccccc}
 Dataset       & Number  & Slope  & Intercept 
	& \multicolumn{3}{c}{Observed scatter ($\sigma$)}
	& $\bar O_c$
	& \multicolumn{3}{c}{Intrinsic scatter ($\sigma$)} \\
     &&&&&lower&upper &&&lower&upper\\
\hline
(1) All members & 275 & $ -0.128 \pm 0.001 $ & $ 3.37 \pm 0.1 $
         & 0.0744 & 0.0658& 0.0835  
         & 0.029 & 0.069 & 0.059 & 0.078\\

(2) $V_{13}<17$ & 175 & $ -0.139 \pm 0.003 $ & $ 3.55 \pm 0.2 $
         & 0.0674 & 0.0585& 0.0751  
         & 0.025 & 0.063 & 0.053 & 0.071\\

(3) All with morph & 129 & $ -0.142 \pm 0.003 $ & $ 3.59 \pm 0.2 $
         & 0.0599 & 0.0506& 0.0675  
         & 0.024 & 0.055 & 0.044 & 0.063\\

(4) E\&S0 morph & 97 & $ -0.137 \pm 0.003 $ & $  3.5 \pm 0.3 $
         & 0.0552 & 0.0471& 0.0638  
         & 0.024 & 0.05 & 0.041 & 0.059\\

(5) S0 morph & 71 & $ -0.125 \pm 0.007 $ & $ 3.32 \pm 0.4 $
         & 0.0583 & 0.0453& 0.0683  
         & 0.023 & 0.053 & 0.039 & 0.064\\

(6) E morph & 26 & $ -0.164 \pm 0.006 $ & $ 3.91 \pm 0.5 $
         & 0.0438 & 0.0298& 0.0504  
         & 0.025 & 0.036 & 0.016 & 0.044\\

(7) Sp morph & 32 & $ -0.166 \pm 0.01 $ & $ 3.96 \pm 0.7 $
         & 0.0778 & 0.0458& 0.122  
         & 0.025 & 0.074 & 0.038 & 0.12\\

(8) E\&S0 center & 78 & $ -0.132 \pm 0.004 $ & $ 3.43 \pm 0.3 $
         & 0.0517 & 0.0435& 0.0584  
         & 0.024 & 0.046 & 0.036 & 0.053\\

(9) E\&S0 SW & 19 & $ -0.15 \pm 0.009 $ & $ 3.71 \pm   2 $
         & 0.0735 & 0.0085& 0.119  
         & 0.024 & 0.069 &    0 & 0.12\\

(10) E\&S0 inner & 55 & $ -0.149 \pm 0.005 $ & $ 3.71 \pm 0.3 $
         & 0.0542 & 0.0432& 0.0636  
         & 0.024 & 0.049 & 0.036 & 0.059\\

(11) E\&S0 outer & 42 & $ -0.124 \pm 0.003 $ & $ 3.28 \pm 0.3 $
         & 0.0412 & 0.0282& 0.0532  
         & 0.024 & 0.033 & 0.014 & 0.047\\

(12) E\&S0 bright & 49 & $ -0.135 \pm 0.007 $ & $ 3.49 \pm 0.6 $
         & 0.0461 & 0.0366& 0.0528  
         & 0.025 & 0.039 & 0.027 & 0.046\\

(13) E\&S0 faint & 49 & $ -0.135 \pm 0.009 $ & $ 3.48 \pm 0.7 $
         & 0.065 & 0.049& 0.081  
         & 0.023 & 0.061 & 0.043 & 0.078\\

(14) members bright & 140 & $ -0.141 \pm 0.005 $ & $ 3.57 \pm 0.3 $
         & 0.0657 & 0.0557& 0.075  
         & 0.024 & 0.061 & 0.05 & 0.071\\

(15) members faint & 136 & $ -0.123 \pm 0.004 $ & $ 3.29 \pm 0.3 $
         & 0.086 & 0.069& 0.107  
         & 0.054 & 0.067 & 0.043 & 0.093\\

\hline
\end{tabular}
\label{tab:regress}    
\end{table*}

\begin{figure*}
\centering
\centerline{\psfig{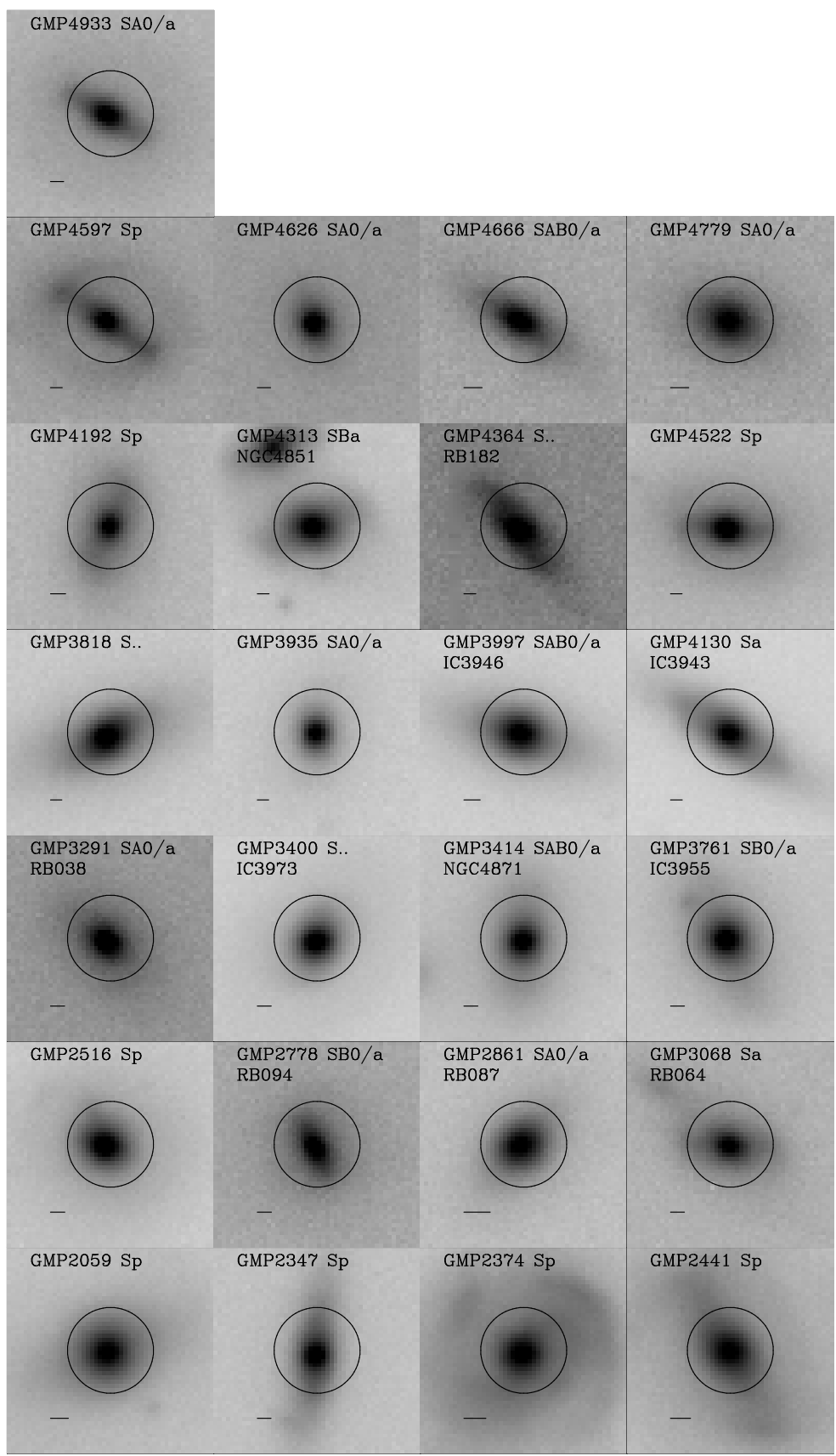}}
\caption{V band images of the late type galaxies which lie within
$3\sigma$ of the CMR for late type galaxies (dataset 7).  The circle
represents the $13''$ diameter aperture used in the measurements of
both the colours and the magnitudes for the results in table
\protect\ref{tab:regress}. The line in the lower left of each panel
shows the width of the seeing disk in each image. The CMR for late types
is identical to that of S0's, further strengthening the argument that
these galaxies have mostly old stellar populations. }
\label{fig:Sp_red}
\end{figure*}
\begin{figure*}
\centering
\centerline{\psfig{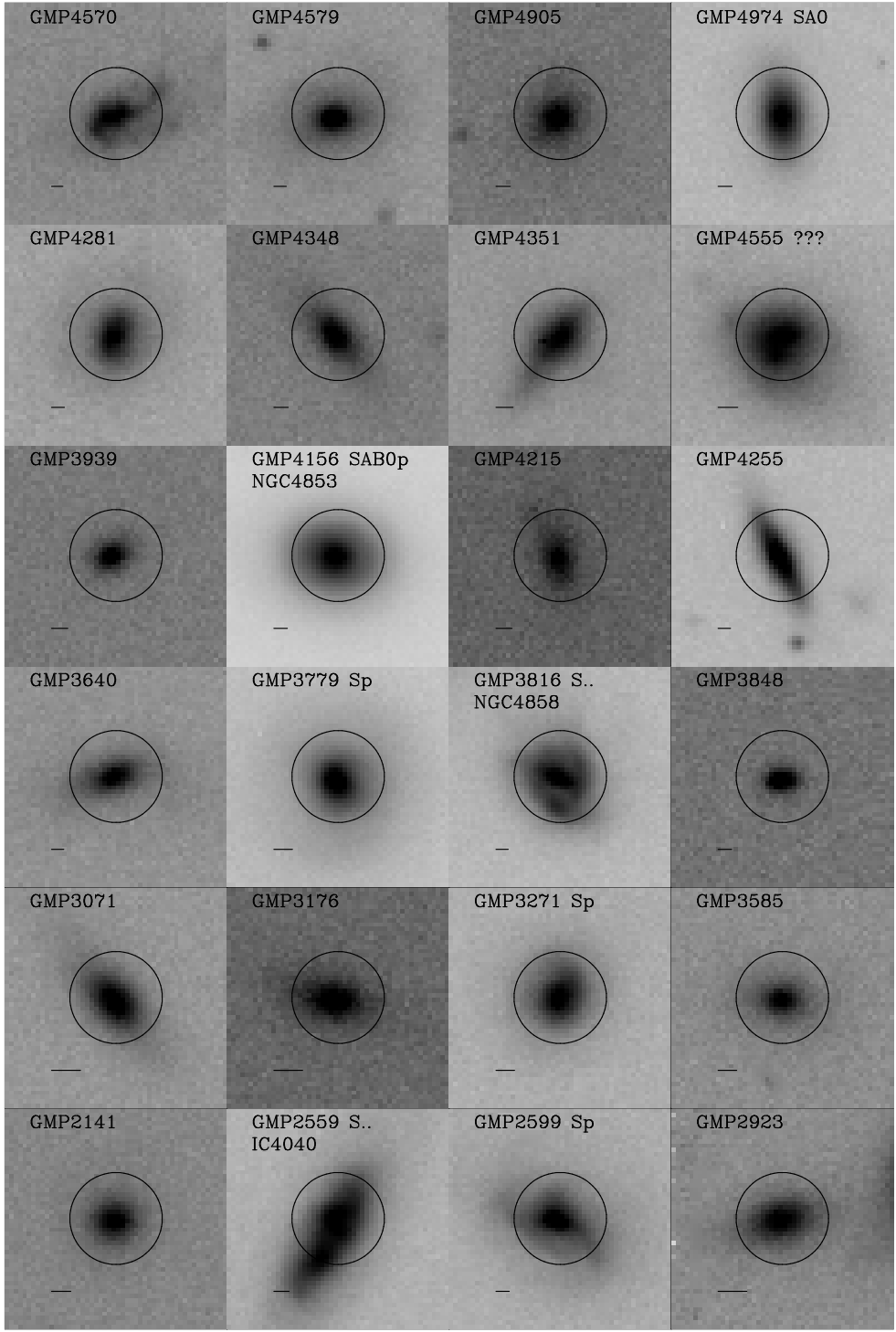}}
\caption{V band images of all of the galaxies which lie more than
$5\sigma$ blue-ward of the CMR for the members dataset (dataset 1). In
addition to the galaxies identified as starburst or post-starburst in
figure~\ref{fig:Sp_blue} (also present in this figure) GMP4255 (D44)
is identified as a post-starburst and GMP4579 (D45) is identified as a
starburst by Caldwell~et~al.~(1993). Text and symbols in the plot are
as for figure \protect\ref{fig:Sp_red}.  }
\label{fig:members_blue}
\end{figure*}

In order to verify whether the findings of the following sections can
be attributed to variations in the CMR, rather than selection biases
in the sample, we have investigated luminosity and morphological
segregation within the samples. Using a K-S test, we find that the
late-type and S0 samples have statistically indistinguishable
luminosity distributions, but that the early type sample is on average
1mag brighter. We also find no correlation between luminosity and
angular distance from NGC4874, nor any significant difference between
the luminosity distribution of the E\&S0 inner and E\&S0 outer
datasets (see figure \ref{fig:E+S0_Lfn}).

\subsection{Morphological dependence of the CMR}
\label{sec:CMR_morph}

In this section we examine variations in the CMR of galaxies of
different morphological types. We use the broad morphological types
defined in table~\ref{tab:MorphNumbers}. The dividing lines
between the various types is somewhat arbitrary, and we err towards
the later morphological types, i.e. we classify a galaxy of type
E/SA0 as S0, and one of type SA0/a as late type. We investigate the
scatter of the main ridge line of the CMR, and the amount of blue-ward
scattering separately. Initially we concentrate on the ridge line.
Looking at the morphologically segregated datasets (3,4,5,6 and 7) in
table~\ref{tab:regress}, they all have levels of intrinsic scatter
indistinguishable within the measurement errors ($\sim0.05$ mag),
except for the elliptical galaxies (dataset 6) which has significantly
lower levels of scatter ($0.036 mag$).

The only exception is the late type galaxy dataset (7): however, even
this data set includes many objects that lie on the CMR ridge line.
Six of the late type galaxies are very blue compared to the CMR
ridge-line. The presence of these blue galaxies is not
surprising. Figure~\ref{fig:Sp_blue} shows snapshots of the V band
images of these galaxies, which even with our poor spatial resolution
can be made out to be very obviously late type. The only S0 galaxy
amongst these, NGC4853, was identified by Caldwell et~al.~(1993) as a
post-starburst galaxy, and is included here with the late types due to
peculiar asymmetries in its light profile. Perhaps the most surprising
aspect of these blue galaxies, is not their presence, but the fact
that there are only seven of them, out of the 32 late types in our
sample. If we apply a $3\sigma$ clipping to this data set, we obtain a
CMR ridge-line that is indistinguishable from the S0
types. Figure~\ref{fig:Sp_red} shows snapshots of the V band images of
these `red' late types. All of the galaxies in figure~\ref{fig:Sp_red}
are within $3\sigma$ of the best fit CMR for dataset 6. Some of the
galaxies are borderline SA0/a, however they are the minority, and
cannot in themselves explain the low scatter. All of the `red' late
type galaxies appear to have fainter disks compared to the `blue' ones
in figure~\ref{fig:Sp_blue}, however statistical tests show no
correlation between a galaxy's Dv and its CMR residual ($(U-V) -
(mV+c)$, where $m$ and $c$ are the best fit slope and intercept for
the CMR). There is however a strong correlation between CMR residual
and colour gradient, and whereas the majority of galaxies have colour
gradients in the sense that they get bluer with increasing radius, the
blue galaxies have a colour gradient in the opposite direction (they
get redder with increasing radius). Thus, although the `blue' galaxies
are not necessarily more compact (c.f. Moss \& Whittle, 2000), the
`blue' light is more concentrated than the CMR galaxies. We conclude
that the `red' late types are `anemic' (e.g. Van~Den~Bergh,~1991),
i.e. that they have lost their $HI$ gas through interactions with the
intra cluster medium.
   
The only other dataset with large numbers of `blue' galaxies, is
dataset 1 (all cluster members). These blue galaxies are, in addition
to the late type galaxies noted above, morphologically untyped. Figure
\ref{fig:members_blue} shows V band images of all of the galaxies
which deviate from the CMR ridge line for dataset 1 by more than
$5\sigma$. It is immediately obvious that they are predominantly of
late type (e.g. GMP4570), although in some cases it is difficult to
tell (e.g. GMP3848).

\subsection{Luminosity dependence of CMR}
\begin{figure*}
\centering
\centerline{\psfig{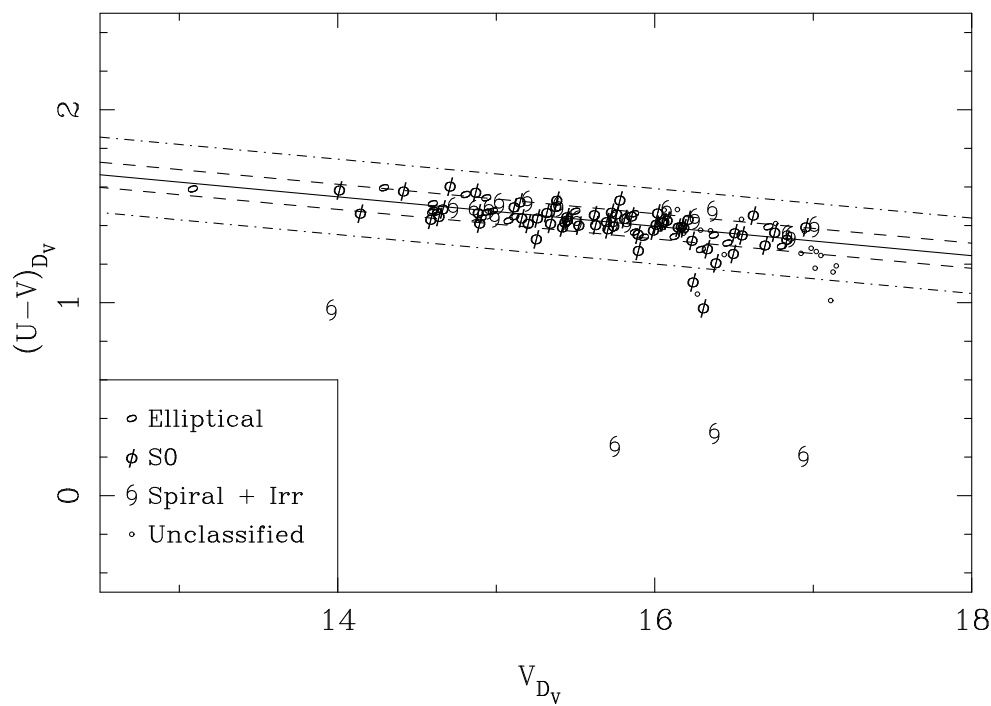}}
\caption{The colour-magnitude relation measured with in the photometric
diameter $D_V$. In this plot, the area within which the photometry is
measure varies according to galaxy size. $V_{D_V}$ is the V-band
magnitude within the $D_V$ diameter; $(U-V)_{D_V}$ is the colour
within this diameter.
}
\label{fig:dvcmr}
\end{figure*}

The last set of datasets are the ones where we segregated the galaxies
according to their luminosity (12,13,14 and 15). We have separated the
early type galaxies dataset (4) into two halves of equal numbers, with
the bright half of the galaxies in dataset 12, and the faint half in
dataset 13. Because dataset 1 spans a much larger range in luminosity
than do the datasets with only morphologically typed galaxies, we
split that into two halves too (datasets 14 and 15).

The bright and faint early type galaxies have indistinguishable slope
and intercept, however the faint sample has greater intrinsic
scatter. This could be related to the increasing numbers of blue
galaxies at fainter magnitudes. As these galaxies start to become
apparent after $V_{13}>16$, they are mainly untyped in our study.

The two halves of the complete members dataset, have differing
slopes. It is possible that this could simply be due to aperture
effects, but as we don't have reliable $D_V$ measurements for the
faint half of the dataset, we cannot check this directly (see below).
Although the observed scatter in the faint sample is larger than that
of the bright sample, it has twice the mean observational error ($\bar
O_c$). Once the difference in the observational errors is taken into
account, both the bright and the faint samples have similar values for
the intrinsic scatter.

A common concern over the interpretation of the CMR as a constraint on
galaxy formation (eg., Kauffmann \& Charlot, 1998; Bower et al., 1998)
is that the role played by colour gradients within galaxies. Is it
possible, for example, that the slope of the CMR is due to the metric
aperture measuring a larger fraction of the total light in small
galaxies than in larger ones?  To address this effect, we have
measured the colour within the $D_V$ diameter of each galaxy. This
gives a measure of the colour of each galaxy that scales with the
properties of the galaxy. This approach is preferable to measuring the
colour within a fixed fraction of the total light since $D_V$ is
(usually) a radius at which the colour can be accurately defined and
does not require extrapolation.  In contrast, measurements based on
the effective radius, $R_e$, require extrapolation of the radial
profile, and often require the colour to be measured at a radius where
the signal to noise ratio is low.

\begin{table}
\caption{Results of regression analysis on the aperture magnitudes
measured within $D_V$ diameter apertures. For a
complete description of each dataset see the main text. The errors
quoted for the slope are $1\sigma$
bootstrap errors. }
\centering{
\begin{tabular}{lcc}
 Dataset       & Number  & Slope  \\
\hline
(3) All with morph & 111 & $ -0.0754 \pm 0.005 $  \\

(4) E\&S0 morph & 86 & $ -0.0819 \pm 0.004 $  \\

(5) S0 morph & 63 & $ -0.082 \pm 0.006 $  \\

(6) E morph & 23 & $ -0.0911 \pm 0.003 $  \\
\hline
\end{tabular}}
\label{tab:regress_Dv}    
\end{table}

The colour-magnitude relation measured within $D_V$ is shown in
Figure~\ref{fig:dvcmr}. The slope of this relation is weaker than that
measured within the fixed diameter, it is still clearly evident.
Table~\ref{tab:regress_Dv} shows the slope of the relation for the
whole sample and for the early-type data sets.  Bower at~al.~(1998)
estimated that the approximately $1/3$ of the CMR slope was due to
aperture effects. Comparison with Table~\ref{tab:regress} shows that
the reduction in the slope is consistent with this. This data-set
therefore confirms that the existence of the CMR slope is not an
artifact due to radial colour gradients in the galaxies. Any model of
cluster galaxy formation must therefore be able to simultaneously
explain the small scatter of the relation and its slope. Importantly,
this constrains the factor by which the masses of cluster galaxies
can grow through random collisions between objects of different
colours (see Bower et~al.~(1998) for a fuller discussion).

\subsection{Environmental dependence of the CMR}

In order to investigate the environmental dependence of the CMR, we
have defined subsamples of the cluster members according to their
position in the sky. Datasets 10 and 11 contain galaxies which are
closer and further than $15$arcmin of NGC4874 respectively, while
datasets 8 and 9 contain galaxies which are either nearer to NGC4874
or NGC4839 respectively (see table~\ref{tab:Cuts}). We have
restricted ourselves to early type galaxies (ellipticals and S0s) to
avoid as much as possible any bias arising from the
morphology--density relation within the cluster, although
section~\ref{sec:CMR_morph} showed there to be little variation in
the properties of the CMR ridge-line between the different
morphological types.

\subsubsection{Significance of the environmental change}
\begin{figure}
\centering
\psfig{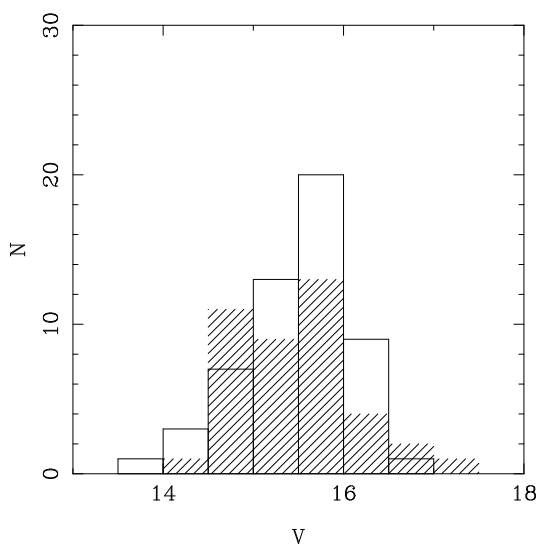}
\caption{ The luminosity distributions for the inner and outer E+S0
datasets (10 and 11). The lined histogram represents the luminosity
distribution for the outer dataset (11) and the outlined histogram
represents the luminosity distribution of the inner dataset (10).
Both distributions are statistically indistinguishable using both a
K-S test, and a Student's t test (see section~\protect\ref{sec:CMRs})
}
\label{fig:E+S0_Lfn}
\end{figure}

 The low number of galaxies in the SW sample make it very difficult to
measure its scatter, the measured scatter in the full dataset is
$0.07\pm0.06$, so we instead concentrate on the scatter in the central
and outer samples (datasets 10 and 11). Here we see something
unexpected. The outer dataset has less scatter (both observed and
intrinsic) than the inner dataset. Although it is only marginally
significant, it is just the opposite of what we would
expect. Figure~\ref{fig:datasets1} (panels 10 and 11) show the CMRs
for these datasets. From them we can see that they both have similar
numbers of elliptical and S0 galaxies. There are too few elliptical
galaxies in each dataset to be able to measure the scatter reliably
for just the ellipticals in each one, but it is possible for the S0
galaxies. This shows the effect to be unrelated to morphology, with
the 41 inner S0 galaxies having an observed scatter of $0.0597 \pm
0.02$ and the 30 outer galaxies having an observed scatter of $0.0377
\pm 0.01$. Figure~\ref{fig:E+S0_Lfn} shows the luminosity distribution
for both datasets, which are statistically indistinguishable using
both a K-S test, and a Student's t test (see section~\ref{sec:CMRs}).

%


%
\begin{figure}
\centering
\centerline{\psfig{figure=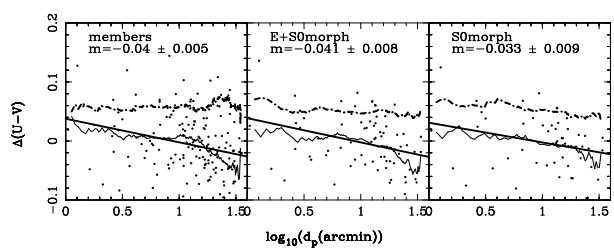,width=8cm,angle=270}}
\caption{ The residuals $(U-V) - (mV + c)$ of each galaxy from the
best fit CM relation plotted against ${\rm log_{10}}$ projected distance
from NGC4874 in arc minutes for three of the datasets described in the
text. Objects around NGC4839 have been excluded. The thin line shows
the running biweight location indicator and the thick line shows the
running scale indicator. Both indicators have a maximum binsize of
40. This reduces to 10 at both ends. The solid straight line is a
biweight fit to all of the data, the slope of which ($m$), and its
$1\sigma$ bootstrap error is shown in each panel.}
\label{fig:runscatter}
\end{figure}

Another approach to analysing the environmental dependence of the CMR
is to analyse the data by averaging over radial bins. Because there is
insufficient data within a single bin, we use a running biweight to
analyse the radial dependence. We have excluded the galaxies around
NGC4839 group in order to ensure that the radial sample also
corresponds to a gradient in density.  We ordered the galaxies
according to their projected distance from NGC4874, and calculated
running biweight location and scale indicators for the residuals from
the CMR ridge line for three datasets, the full dataset (1), the early
type galaxy dataset (4) and a dataset of just the S0 galaxies (5) (see
figure~\ref{fig:runscatter}). In all three panels, the $(U-V)$
residuals are calculated using the best fit CMR to the whole data from
table~\ref{tab:regress}, such that negative values of $\Delta(U-V)$
imply a galaxy is positioned to the blue side of the CMR ridge
line. The first panel, using all the cluster members shows no
increase in scatter with radius, however this could be affected by
greater numbers of late type galaxies in the outer parts of the
cluster. We therefore also show the colour residuals of the early type galaxy
datasets (4 and 5). Again the scatter remains almost constant, with a very
small downward gradient. All three panels show that the residuals
from the CMR seem to getting systematically bluer towards the edges of
the cluster. It seems unlikely that this could be an age effect
without also incurring an increase in the scatter of the CMR, which
leaves two possibilities. Firstly it could simply be a radial drift in
our photometric zero points. Although we checked for this against the
data of BLE92a (see figure~\ref{fig:UVRGBradscat}), it only
extends out to a radius of $15$ arc minutes, which is also where this
effect begins to be noticeable. We can however make a quick estimate
of the expected drift in our photometry. We have on average 20 objects
in the overlap regions, with RMS photometric errors of $0.026$ mag
(see table~\ref{tab:photscat}), which gives an RMS colour error
between images of $0.026/\sqrt{20}$. Now to get to a radius of
$30$arcmin, we need to traverse at least 3 image boundaries, so the
error accumulated is $\sqrt{3} \times 0.026/\sqrt{20} = 0.01$
mag. However, the radius can be calculated in many different
directions, and the photometric zero point for each image was indeed
calculated in an iterative way such that any errors would dissipate in
a two dimensional manner, so the value of $0.01$mag (much smaller than
the value of the colour gradient) can be regarded as an upper limit.

\subsubsection{Dust}
A possible explanation for the gradient could be that the galaxies in
the center of the cluster are being reddened due to intracluster
dust. The upper limit on the reddening through dust in the core of
Coma, as compared to the field is $E(U-V) \leq 0.08$ mag Ferguson
(1993) could account for this amount of reddening, it could also add
an extra source of scatter to the central parts of the cluster not
present in the outer parts. Galaxies behind the cluster would appear
both fainter and redder than an identical galaxy in front of the
cluster. This would tend to increase the scatter in the CMR, but only
in the central parts.

We can estimate the contribution of this dust to the scatter in the
core of the cluster as follows. We define $r$ as the distance from the
core along the line of sight. If $r_v$ is the virial radius of the
cluster, then $r=-r_v$ is the front of the cluster, and $r=r_v$
is the rear of the cluster. We also assume that the galaxies and dust
have the same isothermal density distribution, out to the virial
radius:
\[
\delta(r) = \left\{ \begin{array}{cl}
        {1 \over 1 + \left({r\over r_c}\right)^2} &,\ \ \ |r|\leq r_v\\
	0 &, \ \ \ |r| > r_v \end{array} \right. 
\]
where $r_c$ is the core radius. 
The amount by which a galaxy at $r$ is reddened, is then
\begin{eqnarray}
R(r) &=& {\displaystyle \alpha r_c \left[arctan\left({r\over r_c}\right) +
     arctan\left({r_v\over r_c}\right)\right]}\nonumber
\end{eqnarray}
$\alpha$ is a constant, which we chose in order to satisfy our
boundary condition, that we get the required reddening in the center
of the cluster, i.e. $R(0) = 0.08mag$ \cite{Ferguson93}.
We then calculate the mean and standard deviation of $R(r)$:

\[
\begin{array}{ccccc}
\bar R &=&\displaystyle \alpha r_c arctan\left( {r_v \over r_c}\right)
       &=& 0.08 mag\nonumber
\end{array}
\]
and 
\[
\begin{array}{ccccc}
\bar {R^2} &=& \displaystyle {4\over 3} \left ( \alpha  r_c  arctan \left[
{r_v \over r_c}\right] \right)^2 &=& { 4 \times 0.08 \over 3} mag^2\nonumber
\end{array}
\]
It should be noted that because of our choice of distributions for the
dust and galaxies, both $r_c$ and $r_v$ have cancelled out of the
calculations. The standard deviation of $R(r)$ is then
\[
\begin{array}{ccccc}
SD(R) &=& \sqrt{\bar {R^2} - {\bar{R}}^2} &=& 0.046 mag \nonumber
\end{array}
\]

It should be remembered that this is a very rough model for the
distribution of dust and galaxies in the cluster. The distribution of
dust especially is very poorly known. Clearly from the fact that the
levels of scatter in this model is greater that that measured for
elliptical galaxies, we can say that the dust is either not
distributed as an isothermal sphere, or that the limit for dust in the
core is lower than the number quoted by Ferguson~(1993).

\subsubsection{Age}
The other possibility for the bluing towards the edges, is a
difference in mean galactic age. Using the models for a single burst
stellar population of age 10Gyr from Bower et~al.~(1998), we find
that $d(U-V)/dt \sim 0.03{\rm mag/Gyr}$. This would make the outer
galaxies approximately 2Gyr younger than the central
galaxies. Assuming younger ages for the galaxy population would make
the difference in age between the inside and the outside smaller,
i.e. if the galaxies are only 5Gyr old, the difference in age between
the inner and outer galaxies is only 1Gyr. Abraham et~al.~(1996)
find a $(g-r)$ colour gradient with projected radius in the $z=0.23$
cluster Abell 2390 of $m=-0.08 mag~log10(r_p)^{-1}$, which they
attribute to an age trend. To compare the Coma colour gradient with
that of A2390, we used template early type galaxy spectra to K correct
the Coma colours to the redshift of Abell 2390, and to convert them
from $U-V$ to $g-r$. We find that the gradient shown in figure
\ref{fig:runscatter} for early type galaxies, is transformed into
$m=-0.024 mag~log10(r_p)^{-1}$, a third of that measured in A2390.

A similar argument to the one above for the increased scatter in the
core due to dust also applies in this case. When we look at the core,
we also include galaxies in the foreground and background which are
not in the cluster core, so are bluer than the core galaxies. This
effect is not as large as the dust effect however, because the
galaxies behind the cluster are just as blue as galaxies in front of
it, so the effect is roughly half that expected from the dust model.

\section{Conclusions}

We have placed new limits on the levels of scatter in the $(U,V)$ CMR
of the Coma cluster. The cluster members were split into groups
depending on their morphology, luminosity or position on the sky, and
the CMR was studied in each of them. We found the properties of the
ridge-line to be surprisingly consistent between all of these groups.
We have also calculated upper and lower limits for the intrinsic
scatter in each galaxy sample, taking into account the low number
statistics that we are dealing with for some of them. The results are
presented in tables~\ref{tab:regress}-\ref{tab:regress_Dv}.

We find no variation in the slope of the CMR ridge-line between
Elliptical and S0 morphological types, The late type galaxies in the
cluster have a marginally steeper slope. This could be connected to
the increased blue scatter we find towards the faint end of the
CMR. In the galaxies for which we have morphological types, all of
these very blue galaxies are late types. All of the `blue' galaxies,
even where we don't have morphological data, have colour gradients in
the opposite direction to the normal CMR galaxies, i.e. the blue
galaxies get redder with increasing galactic radius. This could be due
to star formation constrained to the galaxy core and agrees with the
findings of Moss \& Whittle (2000), who show that a disturbed cluster
galaxy morphology is a strong predictor of compact H$\alpha$
emission. Figure~\ref{fig:members_blue} shows that even with our poor
spatial resolution, which tends to make galaxies look of an earlier
type, many of the unclassified blue galaxies are also of late type.

Although the bluest galaxies, tend to have late type morphology, there
are many galaxies with late-type morphology whose colours are
indistinguishable for E and S0 galaxies.  The presence of such a large
fraction of late-type galaxies on the CMR ridge line, with no increase
in the CMR scatter, is surprising.  However, figure~\ref{fig:Sp_red}
shows that our $13$arcsec aperture is dominated by bulge light, and
that in every case, the galaxies possess only a very faint disk. This
could be a low redshift analogue of the trend seen in high redshift
clusters by Dressler et~al.~(1997) who conclude that ellipticals
predate the cluster virialisation, but that late type galaxies turn
into S0 galaxies upon encountering the cluster. These anaemic spirals
are likely to be galaxies in which star formation has been suppressed
by ram-pressure stripping, but which still retain late-type
characteristics (Poggianti et al., 1999).

Using the photometric diameter $D_V$, we have compared the colour
magnitude relation found within a fixed $13''$ diameter and that 
measured within $D_V$. As expected, that measured within $D_V$ is
shallower, due to colour gradients within the galaxies. 
The slope of the relation is still easily seen, however, showing that 
the intrinsic colours of galaxies vary systematically as a function
of magnitude.  

The slope of the CMR also remains constant as a function of radius
within cluster, as expected from a universal metallicity--mass
relation.  We subtract the mean relation from the colours in order to
study the gradients within the cluster.  We find evidence for a
gradient in the CMR corrected colours with projected cluster
distance. Using a naive calculation for the expected slope in the
photometric zero points, we conclude that it is at least a $6\sigma$
result. The slope of the gradient is approximately one third the size
of that found by Abraham et~al.~~(1996) in the $z=0.23$ cluster Abell
2390, who attributed this to a gradient in the mean ages of the
galaxies. We also find some evidence for decreasing scatter in the
early type galaxies towards the outskirts of the cluster as compared
with the central parts. The upper limit for $E(U-V)$ \cite{Ferguson93}
in the cluster core is approximately equal to the $U-V$ colour
gradient observed, however we calculated the increased scatter
produced in the CMR by the presence of enough dust in the cluster core
to account for the colour gradient, and found that it was greater than
the scatter observed in the elliptical galaxies. We therefore conclude
that there cannot be sufficient dust in the cluster core to account
for the entire gradient, and at least some of this gradient must be
due to a systematic variation in galaxy age.

By comparing the colours with the stellar evolution models of Bower
et~al.~(1998), we estimate that the maximum age difference between the
galaxies in the center of the cluster, and in the outskirts is
2Gyr. Although younger galaxies show more scatter in the CMR than old
galaxies (BLE92b; Bower et~al.~1998), the core of the cluster is
contaminated by young galaxies in front of and behind the cluster.\\

\noindent{\bf Acknowledgements}\\ 
We thank John Lucey, Ian Smail and Jim Rose for help and useful
discussions. We would also like to thank the referee for many useful
comments\\

This research has made use of the NASA/IPAC Extragalactic Database
(NED) which is operated by the Jet Propulsion Laboratory, California
Institute of Technology, under contract with the National Aeronautics
and Space Administration.

%
%
%
%

\section*{Appendix A. Seeing corrections}
\label{sec:seeing}

  In this discussion, we restrict ourselves to a circularly symmetric
PSF. All our measured properties are circularly averaged, so any non
spherical symmetry in the PSF, due maybe to poor tracking or focus,
would cause only second order effects \cite{Saglia93}.

 The Fourier transform of the PSF can be predicted using atmospheric
turbulence theory to be $exp[-(kb)^{5/3}]$ \cite{Fried66,Woolf82},
where the scaling parameter $b=FWHM/2.9207006$. We generalise this to
\begin{equation}
\label{eqn:seeingPSF}
\hat p_\gamma(k) = exp\left[-(kb)^\gamma\right]
\end{equation}
where $\gamma$ controls the amount of light in the wings of the
PSF. $\gamma=2$ corresponds to a Gaussian profile, while the
theoretically predicted value of $\gamma=5/3$ gives a more wingy
PSF. Lower values of $\gamma$ produce even larger wings (e.g. Saglia
et~al.,~1993).

\begin{figure}
\centering
\centerline{\psfig{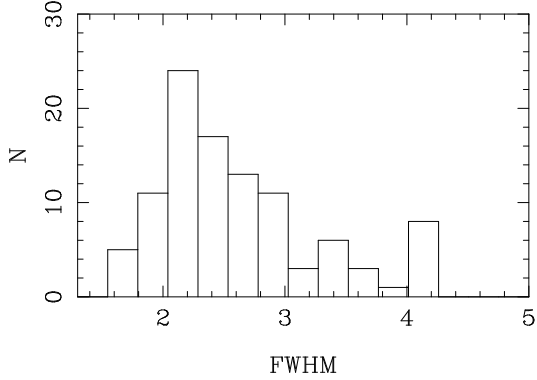}}
\caption{ The distributions of the seeing FWHM obtained by least
squares fitting of the profile of the brightest stars in each image to
the theoretical seeing PSF described in
equation~\protect\ref{eqn:seeingPSF}.}
\label{fig:FWHM}
\end{figure}

We used least square fits of the brightest stellar objects in each
image to obtain both $\gamma$ and $b$ for each exposure.  Figure
\ref{fig:FWHM} shows the distribution of FWHM for all of our
images. Although the values of $\gamma$ vary from 1.2 to 1.8 (although
mainly clustered around 1.45), there is very little variation between
stars in the same exposure, so when calculating the final seeing
correction for a galaxy, we use the $\gamma$ corresponding to the
image the galaxy was measured from

  The Intensity at a radius $R$ from a source ($I(R)$), is then given
by the convolution of the surface brightness distribution of the
object in the sky ($I^s(R)$) with the PSF ($p_\gamma(R)$).
\begin{equation}
\label{eqn:seeconv}
I(R) = I^s(R) \otimes p_\gamma(R)
\end{equation}
 For stellar objects, we simply take the
intensity distribution on the sky to be a delta function, and for
galaxies, we use the canonical deVaucoleurs $R^{1/4}$ law. 
\[ I^s(R) = 
	\left\{ \begin{array}{cl}
        I_e exp\left\{-7.669 \left[\left(
        \frac{R}{R_e}\right)^{1/4}-1\right]\right\} &,\ \ \ {\rm galaxy}\\
	L\delta_D(R) &,\ \ \ {\rm star} \end{array} \right. \]
where $R_e$ is the half light radius, at which $F(R_e)=F(\infty)/2$,
$I_e=I(R_e)$, $\delta_D(R)$ is the Dirac delta and L is the luminosity
of the star. 

 To convert the luminosity distributions of an object in the sky into
the luminosity distribution of the object on the detector we must
convolve with the PSF (equation~\ref{eqn:seeconv}). 

To seeing correct our objects we require the difference, in
magnitudes, of the flux of an object as measured within an aperture of
radius $R$ on our detector ($F(R)$), and its flux as measured within
the same aperture on the sky ($F^s(R)$). To find the flux inside an
aperture of radius R, we simply integrate the required luminosity
distribution,
%
%

 We found the value of $R_e$ to gave little effect on the seeing
correction, for all $R_e \leq 1$ so we use $R_e=5''$ for all of
our galaxies.
 
Numerical integration techniques were used to perform both the
integrations and the convolutions. The results of which, for a variety
of apertures and seeing FWHM are shown in
tables~\ref{tab:gal_seecorr} and~\ref{tab:star_seecorr}. Both
of these tables assume $\gamma=1.47$, the average value for our
observations.

\begin{table*}
\caption{Table of galaxy seeing corrections based on a galaxy with
$R_e=5$arcsec. To obtain the seeing corrected value, the seeing
correction is subtracted from the observed aperture magnitude. We
parameterise the profile of the PSF using $\gamma = 1.47$, the average
$\gamma$ for our observations. Corrections are show for circular
apertures ranging from 4 to 60arcsec diameters. When seeing correcting
our data, we calculated a correction for each image by fitting a FWHM
and $\gamma$ to it from the bright stars.}
\small
\begin{tabular}{ccccccccccccc}
seeing& \multicolumn{12}{c}{\hrulefill\ Aperture diameter (arcsec)\hrulefill}\\
 FWHM&  4   & 6     & 8     & 10    & 13    & 16    & 20    & 25    & 32    & 40    & 50    & 60 \\
\hline
4.50 & 0.909 & 0.564 & 0.376 & 0.268 & 0.176 & 0.126 & 0.087 & 0.060 & 0.039 & 0.027 & 0.018 & 0.014 \\
4.00 & 0.789 & 0.476 & 0.314 & 0.222 & 0.146 & 0.104 & 0.072 & 0.050 & 0.033 & 0.022 & 0.015 & 0.011 \\
3.50 & 0.667 & 0.391 & 0.254 & 0.180 & 0.118 & 0.085 & 0.059 & 0.041 & 0.027 & 0.018 & 0.012 & 0.009 \\
3.00 & 0.542 & 0.309 & 0.200 & 0.141 & 0.093 & 0.067 & 0.046 & 0.032 & 0.021 & 0.014 & 0.010 & 0.007 \\
2.50 & 0.418 & 0.233 & 0.150 & 0.106 & 0.070 & 0.050 & 0.035 & 0.024 & 0.016 & 0.011 & 0.007 & 0.005 \\
2.00 & 0.300 & 0.165 & 0.106 & 0.075 & 0.050 & 0.036 & 0.025 & 0.017 & 0.011 & 0.008 & 0.005 & 0.004 \\
1.50 & 0.193 & 0.106 & 0.068 & 0.049 & 0.032 & 0.023 & 0.016 & 0.011 & 0.007 & 0.005 & 0.003 & 0.002 \\
1.00 & 0.104 & 0.057 & 0.037 & 0.027 & 0.018 & 0.013 & 0.009 & 0.006 & 0.004 & 0.002 & 0.001 & 0.001 \\
\hline
\end{tabular}
\label{tab:gal_seecorr}
\normalsize
\end{table*}
\begin{table*}
\caption{Table of stellar seeing corrections. To obtain the seeing
corrected value, the seeing correction are subtracted from the
observed aperture magnitude. We parameterise the profile of the PSF
using $\gamma = 1.47$, the average $\gamma$ for our
observations. Corrections are show for circular apertures ranging from
4 to 60arcsec diameters. When seeing correcting our data, we
calculated a correction for each image by fitting a FWHM and $\gamma$
to it from the bright stars.}
\small
\begin{tabular}{ccccccccccccc}
seeing& \multicolumn{12}{c}{\hrulefill\ Aperture diameter (arcsec)\hrulefill}\\
 FWHM&  4   & 6     & 8     & 10    & 13    & 16    & 20    & 25    & 32    & 40    & 50    & 60 \\
\hline
4.50 & 1.093 & 0.573 & 0.332 & 0.214 & 0.129 & 0.088 & 0.060 & 0.041 & 0.028 & 0.020 & 0.014 & 0.011 \\
4.00 & 0.922 & 0.461 & 0.263 & 0.170 & 0.104 & 0.072 & 0.049 & 0.034 & 0.023 & 0.016 & 0.012 & 0.009 \\
3.50 & 0.746 & 0.357 & 0.202 & 0.132 & 0.082 & 0.057 & 0.040 & 0.028 & 0.019 & 0.013 & 0.010 & 0.007 \\
3.00 & 0.573 & 0.263 & 0.150 & 0.099 & 0.063 & 0.044 & 0.031 & 0.022 & 0.015 & 0.011 & 0.008 & 0.006 \\
2.50 & 0.408 & 0.184 & 0.107 & 0.072 & 0.046 & 0.033 & 0.023 & 0.016 & 0.011 & 0.008 & 0.006 & 0.004 \\
2.00 & 0.263 & 0.120 & 0.072 & 0.049 & 0.032 & 0.023 & 0.016 & 0.012 & 0.008 & 0.006 & 0.004 & 0.003 \\
1.50 & 0.150 & 0.072 & 0.044 & 0.031 & 0.020 & 0.015 & 0.011 & 0.008 & 0.005 & 0.004 & 0.003 & 0.002 \\
1.00 & 0.072 & 0.037 & 0.023 & 0.016 & 0.011 & 0.008 & 0.006 & 0.004 & 0.003 & 0.002 & 0.001 & 0.001 \\
\hline
\end{tabular}
\label{tab:star_seecorr}
\normalsize
\end{table*}

\section*{Appendix B. Photometric Calibration via Frame Offsets}

The method we used to generate this system is very similar to that
used in Maddox et~al.~(1990) to homogenise the APM galaxy
catalogue. However, it is far simpler due to our detector's better
flat fielding and its linearity.

  To calculate a set of zero-point offsets for each image the regions
of overlap between each pair of images was examined. An object
positioned in an area where frames $i$ and $j$ overlap will have 
magnitudes $m_i$ and $m_j$ as measured from frames $i$ and $j$
respectively. The actual magnitude for this object is $m_0$, and in the
absence of observational errors, these three quantities can be related
to each other thus,
\begin{equation}
\label{eqn:m0def}
m_0 = m_i + C_i = m_j + C_j
\end{equation}
where $C_i$ is the correction applied to the zero point of image $i$.
If we define 
\begin{displaymath}
T_{ij} = m_i - m_j = C_j - C_i
\end{displaymath}
as the overlap difference between $i$ and $j$, then the offset
correction for any image can be calculated from the overlap differences
and offset corrections of any adjacent overlapping image.
\begin{displaymath}
C_{i_j} = C_j + T_{ij}, \forall j\subset i
\end{displaymath}
where $j\subset i$ denotes any pair of images $i$,$j$ with a valid
overlap region, and $C_{i_j}$ denotes the $C_i$ as calculated from
image $j$.  Firstly we construct an observed estimate of the $T_{ij}$.
\begin{displaymath}
T^e_{ij} = { \sum_{n=1..N_{ij}} (m_{n_i} - m_{n_j}) \over N_{ij}}
\end{displaymath}
 where $N_{ij}$ is the number of bright objects in the region of overlap
between images $i$ and $j$, and the $m_{n_i}$ and $m_{n_j}$ are the
measured magnitudes of object $n$ in images $i$ and $j$ respectively.

 Now we can use a weighted mean of the $T^e_{ij}$ to find a value for
the photometric offsets.
\begin{equation}
\label{eqn:jitter}
C_i^{n+1} = \frac{\displaystyle\left( C_i^nW_i + \sum_{i\subset j} (C_j^n +
T_{ij}^e) W_{ij}\right)} {\displaystyle\left( W_i + \sum_{j\subset i}
W_{ij}\right)}
\end{equation}
where the $C_i^n$ are the $C_i$ calculated in iteration $n$, $W_i$ is
the mean of the weights, $W_{ij}$. These weights were chosen to be
proportional to the number of objects used to calculate the
$T^e_{ij}$, and normalised to be in the range zero to one.
%
%
 We chose to use the number in the overlap ($N_{ij}$), rather than the
inverse of the scatter in the $(m_{n_i} - m_{n_j})$. With the low
number of objects present in our overlaps, the scatter is not always
well determined, and can be artificially low for overlaps with low
$N_{ij}$, just the opposite of the required behaviour.

 This iterative method does not constrain the total photometric offset,
it merely ensures the best possible relative photometry of the system
by removing as much of the drift in the zero point between images as
is possible in a self consistent manner. We therefore arbitrarily
re-normalise the $C_i$ after every iteration so that they have zero
mean.

Equation~\ref{eqn:jitter} was iterated to find the best set of
$C_i$. To measure the progress of the iterations, we construct a
measure of the homogeneity of the system after iteration $n$,
\begin{equation}
\label{eqn:jiterr}
E^n = \sum_{i=1..N} \sum_{i\subset j} W_{ij}(T_{ij} + C^n_i - C^n_j)^2
\end{equation}
where $N$ is the total number of images for which we are trying to
ascertain the $C_i$s. We iterate until the rate of change of $E^n$ has
slowed to less than $E^n/1000$ per iteration.

  Obviously much care has to be taken in the measurement of the
$T^e_{ij}$.  We must ensure that effects such as different seeing
conditions on the two overlapping images do not cause any systematic
offsets.  Although seeing conditions are taken into account for each
object when measuring the $m_i$ (see section~\ref{sec:seeing}),
the effect of seeing on objects adjacent to the aperture are not
corrected for i.e.  more light from an adjacent object will enter the
aperture for images with worse seeing. We therefore measured the $m_i$
in various sized apertures, from $8.8''$ to $26''$ diameter, and using
background annuli from $20''$ to $100''$ diameter with $25''$ and
$3.2''$ widths.  If the $T_{ij}$ are well behaved under all measuring
conditions, then we simply take the median value. If they are not well
behaved, we inspect the region more carefully to ascertain the cause,
or we mark the $T_{ij}$ as unreliable.

  To summarise, we can now construct a homogeneous dataset by adding
to each magnitude $m_i$, the offset correction for the image in which
it was measured ($C_i$) to obtain the object's corrected magnitude
$m_0$ (see equation~\ref{eqn:m0def}). This homogeneous photometric
system however is only relative. To set the overall photometric
offset, we use the fixed aperture magnitudes of BLE92a. We have new
observations of all of the BLE92a sample, so using these galaxies we can
define a transformation between our corrected aperture magnitudes, and
theirs. We conclude that our V band response is similar to that of
BLE92a, but the larger value of the U-band colour term indicates that our
U band response is not such a good match, probably due to our
increased sensitivity to the blue portion of the U bandpass.

\label{lastpage}
\end{document}